\documentclass[preprint,aps,prd,floatfix,nofootinbib,11pt]{revtex4-2}
\pdfoutput=1
\usepackage{graphicx}
\usepackage{amsfonts}
\usepackage{nicefrac}
\usepackage{xcolor}
\usepackage{natbib}

\usepackage{graphicx}
\usepackage{amsfonts}
\usepackage{hyperref}

\usepackage{amsmath}
\usepackage{rotating}
\usepackage{amssymb,amsthm}
\usepackage{soul}

\usepackage{xcolor}
\usepackage[normalem]{ulem}

\newcommand\horthop{\widetilde{h}{}}

\usepackage{microtype}

\def \bh {\mbox{{\bf h}}}

\begin{document}


\title{Teleparallel Robertson-Walker geometries and applications}

\author{A. A. Coley}
\email{aac@mathstat.dal.ca}
\affiliation{Department of Mathematics and Statistics, Dalhousie University, Halifax, Nova Scotia, Canada, B3H 3J5}

\author {A. Landry}
\email{a.landry@dal.ca}
\affiliation{Department of Mathematics and Statistics, Dalhousie University, Halifax, Nova Scotia, Canada, B3H 3J5}

\author {F Gholami}
\email{ft774404@dal.ca}
\affiliation{Department of Mathematics and Statistics, Dalhousie University, Halifax, Nova Scotia, Canada, B3H 3J5}

\begin{abstract}

\vspace*{0.5cm}

In teleparallel geometries the coframe and corresponding spin-connection are the principal geometric objects and consequently the appropriate definition of a symmetry is that of an affine symmetry. The set of invariant coframes and their corresponding spin connections that respect the full six dimensional Lie algebra of Robertson-Walker affine symmetries are displayed and discussed. We will refer to such geometries as teleparallel Robertson-Walker (TRW) geometries, where the corresponding derived metric is of Robertson-Walker form and is characterized by the parameter  $k = (-1,0,1)$. The field equations are explicitly presented for the $F(T)$ class of teleparallel TRW spacetimes. We are primarily interested in investigating the $k \neq 0$ TRW models. After first studying the $k=0$ models and, in particular, writing their governing field equations in an appropriate form, we then study their late time stability with respect to perturbations in $k$ in both the cases of a vanishing and non-vanishing effective cosmological constant term. As an illustration we consider both quadratic $F(T)$ theories and power-law solutions.



\end{abstract}

\maketitle

\newpage


\section{Introduction}

In the covariant approach to teleparallel theories of gravity \cite{Krssak_Saridakis2015,Krssak:2018ywd}, the  (co)frame and corresponding spin-connection are the main geometric objects of study (and the metric is a derived quantity). The appropriate definition of a symmetry in a teleparallel geometry is that of an affine symmetry. In 
a teleparallel manifold an affine frame symmetry on the frame bundle is a diffeomorphism from the manifold to itself  that leaves the spin-connection invariant and affects the invariant frame in a particular way, and is defined by the existence of a vector field, ${\bf X}$, which satisfies \cite{Coley:2019zld}:
\begin{equation}
\mathcal{L}_{{\bf X}} \bh_a = \lambda_a^{~b} \,\bh_b \mbox{ and } \mathcal{L}_{{\bf X}} \omega^a_{~bc} = 0, \label{Intro:FS2}
\end{equation}
where $\omega^a_{~bc}$ represents the spin-connection defined with respect to the geometrically preferred invariant frame $\bh_a$ determined by the Cartan-Karlhede algorithm and $\lambda_a^{~b}$ is an element of the linear isotropy group thereby obtained. This definition is {\it a frame-dependent} analogue of the definition of a symmetry presented in \cite{HJKP2018,HJKP2018a}.\footnote{In the Cartan-Karlhede algorithm adopted for teleparallel geometry \cite{Coley:2019zld}, the parameters of the Lorentz frame transformations are fixed by the normalization of the components of the  torsion tensor and its covariant derivatives in an invariant manner. This algorithm consequently leads to an invariantly defined frame up to linear isotropy (defined as the (sub)group of Lorentz transformations that leave the torsion tensor and its covariant derivatives invariant). In particular,  the Cartan-Karlhede algorithm consequently provides a set of Cartan invariants which can then be used to uniquely characterize a geometry locally and determine the dimension of the affine symmetry group. }


Using this algorithm, the invariant coframes and the corresponding spin connection that respects the imposed affine frame symmetries were constructed.  In particular, the teleparallel geometries which are invariant under the full $G_6$ Lie algebra of affine symmetries were displayed \cite{preprint}. In addition, the proper coframe has also been obtained in each case and the field equations (FE) displayed. We denote such geometries as teleparallel Robertson-Walker (TRW) geometries \cite{preprint}, where the associated (derived) metric is of Robertson-Walker form  characterized by a constant parameter $k = (-1,0,1)$. We note that this parameter $k$, which is usually interpreted as the constant spatial curvature in the RW pseudo-Riemannian metric, cannot be so related here since the Riemann tensor is identically zero in teleparallel spacetimes; while “k” may be considered as the curvature of the $3$-space, in $4$-space it actually occurs as part of the torsion scalar. We also comment that in the TRW geometries, an appropriate spin connection/coframe pair results in a situation in which the antisymmetric part of the FEs are identically zero.

\newpage

Most of the research up until now has been concerned with the analysis of the flat ($k=0$) TRW cosmological models \cite{Bahamonde:2021gfp,Cai_2015} (and references within). In  particular, specific forms for $F(T)$ have been investigated (using a priori ansatz such as, e.g.,  polynomial functions), and reconstruction methods have been explored extensively (in which the function $F(T)$ is  reconstructed from the underlying assumptions of the models; e.g., a functional form for the solution such as  the scale factor is a simple power-law). Dynamical systems methods (e.g., fixed point and stability analysis) in flat TRW models have been widely utilized \cite{Bahamonde:2021gfp,coley03} (also see \cite{BahamondeBohmer,Kofinas,BohmerJensko,aldrovandi2003}), which include the study of the stability of the standard de Sitter fixed point.


There has been much confusion in the analysis of TRW geometries since it is often unclear whether the  frame and the corresponding spin connection admits the full $G_6$ Lie algebra of spatially homogeneous and isotropic affine symmetries (which are then necessarily isometries of the metric). In particular, for non-zero $k$ the geometries investigated often do not have a $G_6$ of affine frame symmetries (usually only $3$ of the Killing vectors (KVs) considered are affine frame symmetries). Indeed, for the $k=\pm 1$ cases, there have been a number of incorrect attempts to find solutions, often involving the use of complex tetrads or inappropriate spin connections. However, recently frame/connection pairs for geometries satisfying a $G_6$ group of symmetries have been constructed  \cite{Hohmann:2015pva,HJKP2018,HJKP2018a,Hohmann:2018rwf} in a similar (but not equivalent) fashion to that presented in \cite{Coley:2019zld} and utilized here. In addition, when a proper frame is assumed additional confusion can occur.


The earliest attempts to include a non zero $k$ properly were done in \cite{Ferraro,Ferraro2,HJKP2018,HJKP2018a,Hohmann:2018rwf}, although the derivation was questionable in one of the first times the FE were written down \cite{Hohmann2008}. However, it is still not clear within this alternative approach (see also \cite{Hohmann:2015pva,Hohmann2021,Ambrosio,Ambrosio2}) which functions are essential in finding a solution and which depend on the coordinates chosen. In addition, subclasses of geometries with an additional symmetry can't be determined explicitly using this approach; larger symmetry groups must be assumed from the outset and it must then be determined whether they exist or not  by trial and error. Nevertheless, the resulting eqns. are the same as those obtained in \cite{preprint} and displayed below.


We are primarily interested in studying the $k \neq 0$ TRW models here. Geometries with non-zero $k$ have been recently studied in bounce models \cite{bounce} and in
inflation \cite{Capozz} (although the analysis may only apply for $k=1$, due to the use of a complex tetrad). Also, perturbations have been studied in non-flat cosmology \cite{inflat}.

\section{ Teleparallel Robertson-Walker (TRW) spacetimes}

We shall work in coordinates $(t, r, \theta, \phi)$, where the $G_6$ Lie algebra of affine symmetries are \cite{Coley:2019zld}:
\begin{eqnarray} & X_z = \partial_{\phi},~X_y = - \cos \phi \partial_{\theta} + \frac{\sin \phi}{\tan \theta} \partial_{\phi}, X_x = \sin \phi \partial_{\theta} + \frac{\cos \phi}{\tan \theta} \partial_{\phi}, \nonumber \\
& X_1 = \chi \sin \theta \cos \phi \partial_r + \frac{\chi}{r}\cos \theta \cos \phi \partial_{\theta} - \frac{\chi \sin \phi}{r \sin \theta} \partial_{\phi}, \nonumber \\
& X_2 = \chi \sin \theta \sin \phi \partial_r + \frac{\chi}{r} \cos \theta \sin \phi \partial_{\theta} + \frac{\chi \cos \phi}{r \sin \theta} \partial_{\phi}, \label{G6 generators} \\
& X_3 = \chi \cos \theta \partial_r - \frac{\chi}{r} \sin \theta \partial_{\theta}, \nonumber
\end{eqnarray}
where $\chi \equiv \sqrt{1 - kr^2}$. We write $\{ X_I\}_{I=1}^6 = \{ X_1, X_2, X_3, X_x, X_y, X_z\}$. The largest linear isotropy group allowed by a spatially homogeneous geometry is $SO(3)$ and a matrix basis for its Lie algebra is of the form:
\begin{equation}
 \lambda_{\hat{1}} = \left[ \begin{array}{cccc} 0 & 0 & 0 & 0 \\
0 & 0 & 0 & 0 \\ 0 & 0 & 0 & 1 \\ 0 & 0 & -1 & 0 \end{array} \right],~ \lambda_{\hat{2}} = -\left[ \begin{array}{cccc} 0 & 0 & 0 & 0 \\
0 & 0 & 1 & 0 \\ 0 & -1 & 0 & 0 \\ 0 & 0 & 0 & 0 \end{array} \right] ,~
\lambda_{\hat{3}} =  -\left[ \begin{array}{cccc} 0 & 0 & 0 & 0 \\
0 & 0 & 0 & 1 \\ 0 & 0 & 0 & 0 \\ 0 & -1 & 0 & 0 \end{array} \right]  \label{so3LA}
\end{equation}
\noindent We also work with the following coframe:
\begin{equation}
h^a_{~\mu} = \left[ \begin{array}{cccc} 1 & 0 & 0 & 0 \\ 0 & \frac{a(t)}{\sqrt{1-kr^2}} & 0 & 0 \\ 0 & 0 & a(t) r & 0 \\ 0 & 0 & 0 & a(t) r \sin(\theta) \end{array} \right]. \label{VB:FLRW}
\end{equation}

\noindent The derived metric is the Robertson-Walker (RW) metric:
\begin{align}\label{RWspacetime}
ds^2=-dt^2+a^2(t)\left[\frac{dr^2}{1-k\,r^2}+r^2\,\left(d\theta^2+\sin^2\theta\,d\phi^2\right)\right] ,
\end{align}
where the parameter $k = (-1,0,1)$ but it is not interpreted as a constant spatial curvature here. It is this metric that leads to the Friedmann-Lemaitre-RW models in GR.

Assuming that the connection is metric compatible, we necessarily have that $\omega_{abc} = -\omega_{bac}$. The solution to
eqns \eqref{Intro:FS2} for each of the affine symmetry vector fields in \eqref{G6 generators} gives the non-trivial components 
\cite{preprint}:
\begin{eqnarray}
 & \omega_{122} = \omega_{133} = \omega_{144} =  W_1(t), \nonumber \\
& \omega_{234} = -\omega_{243} = \omega_{342} = W_2(t), \nonumber \\
& \omega_{233} = \omega_{244} = - \frac{\sqrt{1-kr^2}}{a(t)r}, \label{Con:FLRW}\\
& \omega_{344} =  \frac{\cos(\theta)}{a(t) r \sin(\theta)} ,  \nonumber
\end{eqnarray}
where $W_1$ and $W_2$ are arbitrary functions. 
This is the general connection for any Riemann-Cartan geometry admitting the symmetry group with generators given by eqn \eqref{G6 generators}. The tensor-part of the torsion tensor is then identically zero and the torsion tensor can be further decomposed into a vector-part, $V_a = T^b_{~ba}$, and a axial-part, $A_a = \epsilon_{abcd} T^{bcd}$.

In order to determine the connections that represent a teleparallel geometry we must then impose the Riemmann flatness conditions, which have the distinct solutions \cite{preprint}:
\begin{itemize}
\item $W_1(t) = 0$, $\displaystyle W_2(t) = \pm \frac{\sqrt{k}}{a(t)}$ where $\displaystyle {\bf V} = - \frac{3\dot{a}(t)}{a(t)} \bh^1, \mbox{\ \rm and\ } {\bf A} = \mp \frac{2\sqrt{k}}{a(t)} \bh^1$,

\item $\displaystyle W_1(t) = \pm \frac{\sqrt{-k}}{a(t)}$, $W_2(t) = 0$ where $\displaystyle{\bf V} = \frac{3(\pm\sqrt{-k} + \dot{a}(t))}{a(t)} \bh^1,\mbox{\ \rm and\ } {\bf A} = 0 $,
\end{itemize}

\noindent where $a=a(t)$ is the frame function and $\dot{a}=\dot{a}(t)$ is its time derivative. Each case above contains the subcase $k=0$. In the first case the $\pm$ solutions are equivalent under a local Lorentz transformation and in the second case the $\pm$ solutions are equivalent under a discrete coordinate transformation $t\,\rightarrow\,-t$.

In conclusion, each of the TRW geometries given by the coframe \eqref{VB:FLRW} with the connection \eqref{Con:FLRW} (and such that the two arbitrary functions $W_1$ and $W_2$ satisfy one of the forms above) is a teleparallel geometry admitting the required symmetry group with generators given by eqn \eqref{G6 generators}.

The torsion scalar for each of  $k=(-1,0,1)$ is of the form
\begin{eqnarray}
T(t) &=& 6\left(\frac{\dot{a}}{a}\right)^2 - 12W_1\frac{\dot{a}}{a}-6W_1^{\,2}+6W_2^{\,2}
\nonumber\\
      &=& 6\left(\frac{\dot{a}}{a}+W_1+W_2\right)\left(\frac{\dot{a}}{a}+W_1-W_2\right)
\nonumber\\
      &=& \frac{2}{3}V^2 - \frac{3}{2}A^2 ,
\end{eqnarray}
where the vectorial and axial term  magnitudes are given by:
\begin{equation}
V^2=-9\left(\frac{\dot{a}}{a}+W_1\right)^2, \qquad A^2 = -4W_2^2.
\end{equation}
We note that a non-trivial axial part of the torsion scalar only occurs in the $k=+1$ case. Also note that there are $4$ independent (Cartan) invariant scalars in the positive $k$ case, while there are only $3$ in the $k=-1$ case \cite{Coley:2019zld}.

\subsection{TRW Proper Frames}

In teleparallel geometry, there always exists a frame in which the spin-connection is zero. This frame is called the proper frame (or Weitzenbock frame). If we assume an orthonormal diagonal coframe in the spherical coordinate system \eqref{VB:FLRW} with corresponding spin connection computed above, we can investigate each of the 3 values for $k$ independently. For $k=-1$ we have $[W_1(t),W_2(t)] = [\delta\sqrt{-k}/a(t),0]$, for $k=0$ we have $[W_1(t),W_2(t)] = [0,0]$, and for $k=+1$ we have that $[W_1(t),W_2(t)] = [0, \delta\sqrt{k}/a(t)]$,  where $\delta=\pm 1$ is a discrete parameter. 
The spin connection can then be determined, for 
a matrix $\Lambda^a_{~b}\in SO(1,3)$, 
from the differential equation
\begin{equation}
\omega^a_{~b}=(\Lambda^{-1})^a_{~c}d\Lambda^c_{~b};\label{DE_SPIN}
\end{equation}
that is, we now simply need to solve this differential equation for $\Lambda^a_{~b}$ for each of  $k=-1$ and $k=+1$ (where $k=0$ constitutes a subcase of both), and hence easily determine a proper coframe $\horthop^a = \Lambda^a_{~b}h^b$, where $h^b$ is given by eqn. \eqref{VB:FLRW}.

\subsubsection{Negative $k$-parameter TRW Case:}

When $k=-1$, a Lorentz transformation that satisfies the differential eqn. \eqref{DE_SPIN} is
\begin{equation}
\Lambda^a_{~b} = \left[\arraycolsep=3pt\def\arraystretch{1.5} \begin{array}{cccc}
    \sqrt{1-kr^2}&                  -\delta\sqrt{-k} r&                                  0&                             0\\
    -\delta\sqrt{-k} r\sin(\theta)\sin(\phi)&   \sqrt{1-kr^2}\sin(\theta)\sin(\phi)&    \cos(\theta)\sin(\phi)&          \cos(\phi)\\
     -\delta\sqrt{-k} r \sin(\theta)\cos(\phi)&   \sqrt{1-kr^2}\sin(\theta)\cos(\phi)&    \cos(\theta)\cos(\phi)&         -\sin(\phi)\\
      \delta\sqrt{-k} r \cos(\theta)&           -\sqrt{1-kr^2}\cos(\theta)&             \sin(\theta)&                   0
\end{array}\right].
\end{equation}
Clearly any global Lorentz transformation multiplying this transformation also gives rise to a solution. Therefore, we can either formulate the $k=-1$ spacetime geometry by using the proper frame $\horthop^a = \Lambda^a_{~b}h^b$ with this Lorentz transformation and with $h^b$ given by eqn. \eqref{VB:FLRW}, which necessarily has a trivial spin connection, or by
using the diagonal coframe \eqref{VB:FLRW} and the corresponding spin connection one-form \cite{preprint}:
\begin{equation}
\omega^a_{~b} =\left[\arraycolsep=3pt\def\arraystretch{1.5} \begin{array}{cccc}
 0                                & -\frac{\delta\sqrt{-k}}{\sqrt{1-kr^2}}dr   & -\delta\sqrt{-k} r d\theta          & -\delta\sqrt{-k} r\sin(\theta) d\phi \\
 -\frac{\delta\sqrt{-k}}{\sqrt{1-kr^2}}dr   & 0                                & -\sqrt{1-kr^2}d\theta       & -\sqrt{1-kr^2}\sin(\theta)d\phi \\
 -\delta\sqrt{-k} r d\theta                & \sqrt{1-kr^2}d\theta              & 0                          & -\cos(\theta)d\phi \\
 -\delta\sqrt{-k} r\sin(\theta) d\phi      & \sqrt{1-kr^2}\sin(\theta)d\phi    & \cos(\theta)d\phi         & 0
\end{array}\right].\label{conn_neg}
\end{equation}

\subsubsection{Positive $k$-parameter TRW Case:}

When $k=+1$, a Lorentz transformation that satisfies the differential eqn. \eqref{DE_SPIN} is
\small
\begin{eqnarray}
& &\Lambda^a_{~b}=
\nonumber\\
&& \left[\arraycolsep=3pt\def\arraystretch{1.5} \begin{array}{cccc}
    1&                  0       &                                  0                 &                 0\\
    0&   \cos(\theta)           &    -\sqrt{1-kr^2}\sin(\theta)                       &     -\delta\sqrt{k}r\sin(\theta)\\
    0&   \sin(\theta)\cos(\phi) &     \sqrt{1-kr^2}\cos(\theta)\cos(\phi)-\delta\sqrt{k}r\sin(\phi) &     \delta\sqrt{k}r\cos(\theta)\cos(\phi)-\sqrt{1-kr^2}\sin(\phi)\\
    0&   \sin(\theta)\sin(\phi) &     \sqrt{1-kr^2}\cos(\theta)\sin(\phi)+\delta\sqrt{k}r\cos(\phi) &     \delta\sqrt{k}r\cos(\theta)\sin(\phi)+\sqrt{1-kr^2}\cos(\phi)
\end{array}\right].
\nonumber\\
\end{eqnarray}
\normalsize
Any global Lorentz transformation times this transformation is again a solution. We can properly formulate the $k=+1$ geometry by constructing the proper frame $\horthop^a = \Lambda^a_{~b}h^b$ using this Lorentz transformation and with $h^b$ given by eqn. \eqref{VB:FLRW}, which necessarily has a trivial spin connection, or by using the diagonal coframe \eqref{VB:FLRW} and corresponding spin connection one-form \cite{preprint}:
\small
\begin{eqnarray}
& &\omega^a_{~b} =
\nonumber\\
&&\left[\arraycolsep=3pt\def\arraystretch{1.5} \begin{array}{cccc}
 0   & 0                                        & 0                                          & 0 \\
 0   & 0                                        &  -\sqrt{1-kr^2}d\theta +\delta\sqrt{k}r\sin(\theta)d\phi        & -\delta\sqrt{k}rd\theta -\sqrt{1-kr^2}\sin(\theta)d\phi \\
 0   & \sqrt{1-kr^2}d\theta -\delta\sqrt{k}r\sin(\theta)d\phi        & 0                                          & \frac{\delta\sqrt{k}}{\sqrt{1-kr^2}}dr-\cos(\theta)d\phi \\
 0   & \delta\sqrt{k}rd\theta +\sqrt{1-kr^2}\sin(\theta)d\phi & -\frac{\delta\sqrt{k}}{\sqrt{1-kr^2}}dr+\cos(\theta)d\phi & 0
\end{array}\right]. \label{conn_pos}
\nonumber\\
\end{eqnarray} \normalsize

\subsection{Field equations}

Using either the proper coframe (and trivial connection) or the diagonal coframe/ connection pair \eqref{VB:FLRW} and \eqref{conn_neg}, the torsion scalar for $k=-1$ is:
\begin{equation}
T=6\left( \frac{\dot{a}}{a}+ \frac{\delta\sqrt{-k}}{a}\right)^2. \label{15}
\end{equation}
We shall assume an energy momentum tensor of the form $T_{ab}=\rho(t)u_a u_b +(u_a u_b+g_{ab})p(t)$, which formally represents a perfect fluid source with energy density  and pressure denoted by $\rho(t)$ and $p(t)$, respectively. The antisymmetric part of the FEs are then automatically satisfied (and hence all of the frames defined above are "good" frames in the terminology of \cite{Krssak_Saridakis2015}) in the $F(T)$ teleparallel geometries. The remaining linearly independent eqns. from the symmetric part of the FEs in the $F(T)$ teleparallel geometries are then given by:
\begin{subequations}
\begin{eqnarray}
-\frac{F(T)}{2}+6F'(T)\left(\frac{\dot{a}}{a}\right)\left(\frac{\dot{a}}{a}+\frac{\delta\sqrt{-k}}{a}\right)&=&\kappa\rho,
\\
 F(T)-6F'(T)\left(\frac{\ddot{a}}{a}+\left(\frac{\dot{a}}{a}+\frac{\delta\sqrt{-k}}{a}\right)^2 \right)-6F''(T)\dot{T}\left(\frac{\dot{a}}{a}+\frac{\delta\sqrt{-k}}{a}\right)&=&\kappa(\rho+3p).
\nonumber\\
\end{eqnarray}
\end{subequations}
Obviously, these coframe/connection pairs also apply in the subcase $k=0$. Care must be taken in the case $k=+1$ since complex valued coframes or spin connections may occur.

Using either the proper coframe (and trivial spin connection) or the diagonal coframe/ connection pair \eqref{VB:FLRW} and \eqref{conn_pos}, the torsion scalar for $k=+1$ is:
\begin{equation}
T=6\left[ \left(\frac{\dot{a}}{a}\right)^2- \frac{k}{a^2}\right], \label{16}
\end{equation}
which is independent of $\delta$.
Assuming a perfect fluid as in the $k=-1$ case, the antisymmetric part of the FEs are identically satisfied and the symmetric part of the FEs are
\begin{subequations}
\begin{eqnarray}
-\frac{F(T)}{2}+6F'(T)\left(\frac{\dot{a}}{a}\right)^2&=&\kappa\rho,
\\
F(T)-6F'(T)\left(\frac{\ddot{a}}{a}+\left(\frac{\dot{a}}{a}\right)^2-\frac{k}{a^2}
\right)-6F''(T)\dot{T}\left(\frac{\dot{a}}{a}\right)&=&\kappa(\rho+3p).
\end{eqnarray}
\end{subequations}
In this $k=+1$ case, there is no dependence on the discrete parameter  $\delta$. Further, the equations once again reduce to the $k=0$ FEs by setting $k=0$.

The energy-momentum conservation equation
\begin{equation}
{\dot{\rho}}  + 3\frac{\dot{a}}{a}(\rho + p) = 0,\label{17}
\end{equation}
follows from the two FEs above.

Notice that the FEs reduce to those of FLRW models when $F(T)=T$.

\subsection{The function $F(T)$}

A number of particular examples of $F(T)$ theories studied in the literature include polynomial functions in $T$, and especially quadratic  $T^2$ theory 
\cite{Bahamonde:2021gfp}. Recently, it has been shown that the theory
\begin{equation}
F(T) = -\Lambda+T+\gamma\,T^2 \label{quad}
\end{equation}
can alleviate a variety of cosmological tensions \cite{saridakis}. It is also of interest to study the theory with $F(T) = T + \gamma T^{\beta}$ or an exponential function.

\section{Analysis and Equations}

\begin{itemize}

\item We recall that if $T=$ const., then the theory reduces to a renormalized GR and we obtain teleparallel analogues of special solutions in GR \cite{Krssak:2018ywd}. Such solutions with a non-vanishing "effective" cosmological constant will be asymptotic to ($k=0$) teleparallel analogues of de Sitter solutions \cite{preprint}. We are only interested here in non-GR type solutions.

\item We shall assume an eqn. of state of the form: $p= \alpha \rho$, with $-1< \alpha \leq 1$, so that $\rho$ is monotonically decreasing
for $H \equiv \frac{\dot{a}}{a} >0$. We are particularly interested in the case $k\neq 0$.

\item A detailed analysis of the theory $F(T) = -\Lambda + T + \gamma T^{\beta}$ or for power-laws solutions is possible.

\end{itemize}

\subsection{Equation for $k=0$}

In the case $k=0$, the non constant torsion scalar is given by $T=6H^2$ and is positive (and we assume that $H \geq 0$). The conservation eqn. is given by (\ref{17}), and so for $p= \alpha \rho$, with $\alpha > -1$, $\rho$ is monotonically decreasing to zero at late times. Note that in the case $k=0$ at the present time $\rho_c\,=\,\frac{3H^2}{8\pi G}\,\sim\,10^{-26}$ $kg/m^3$. For positive $\rho$ we have that $-\frac{F}{2} + T\,F^{\prime}  > 0$. From the two FEs we then find that (compare with \cite{BohmerJensko}):
\begin{equation}
\frac{dT}{d\tau} =  \frac{3(1+\alpha)(\frac{F}{2} - T\,F^{\prime} )}{ (\frac{F^{\prime}}{2}+ T\,F^{\prime\prime} )} \equiv \mathcal{F}(T),
\label{18}
\end{equation}
where the new time coordinate $\tau$ is defined by 
$$\frac{d\tau}{dt} \equiv H \equiv \sqrt{\frac{T}{6}},$$
which is well defined since $H>0$. Note that for positive $\rho$, $(\frac{F}{2} - T\,F^\prime ) < 0$ and so
$\mathcal{F}(T) < 0$ and $\frac{dT}{d\tau} < 0$ if $(\frac{F^{\prime}}{2} + T\,F^{\prime\prime} ) > 0$. From \eqref{18} we have that:
\begin{equation}
\frac{1}{\mathcal{F}}=-\frac{1}{3(1+\alpha)}\frac{d}{dT} \left[\ln\left(\frac{F}{2} -T\,F^{\prime}\right)\right] ,     
\label{18a}
\end{equation}
so that eqn. \eqref{18} can be integrated to obtain
\begin{eqnarray}
\frac{F}{2} -T\,F^\prime &=& -\tilde{A}_0 e^{-3(1+\alpha)\tau},
\label{18b}
\end{eqnarray}
where $\tilde{A}_0$ is a positive constant.

\subsection{Equation for $k=\pm 1$}


When $k=-1$, $T$ is given by (\ref{15}), and we obtain (where here $\frac{d\tau}{dt} \equiv \sqrt{\frac{T}{6}}$):
\begin{eqnarray}\label{101}
\frac{dT}{d\tau} &=&\,\mathcal{F}(T)\,\sqrt{1-\frac{6\,K}{T}}\,\left[\frac{1+\left(\frac{2\left(1 +3 \alpha\right)\,T\,F'(T)}{\left(1 + \alpha\right)\left(\frac{F(T)}{2}-T\,F'(T)\right)}\right)\,\frac{K}{T}}{1-\left(\frac{6\,T\,F''(T)}{\left(\frac{F'(T)}{2}+T\,F''(T)\right)}\right)\,\frac{K}{T}}\right] ,
\end{eqnarray}
\normalsize
where $K=K(T) \equiv \frac{\sqrt{-k}}{a^2}$. For $\frac{K}{T}\ll 1$, we can approximate this as:
\begin{equation}\label{102}
\frac{dT}{d\tau} \approx \,\mathcal{F}(T)\,+\,\mathcal{G}(T)\,K(T) ,
\end{equation}
where 
\begin{equation}\label{103}
\mathcal{G}(T)= \frac{6\left(1 +3 \alpha\right)\,F'(T)}{\left[\frac{F'(T)}{2}\,+T\,F''(T)\right]}+\frac{9\left(1 + \alpha\right)\left[\frac{F(T)}{2}-T\,F'(T)\right]\left[T\,F''(T)-\frac{F'(T)}{2}\right]}{T\,\left[\frac{F'(T)}{2}\,+T\,F''(T)\right]^2} .
\end{equation}

When $k=+1$, $T$ is given by (\ref{16}) (where $\frac{d\tau}{dt} \equiv H$), we obtain:
\begin{eqnarray}\label{104}
\frac{dT}{d\tau} &=&\,\mathcal{F}(T)\,\sqrt{1+\frac{6\,K}{T}}\,\left[\frac{1-\left[\frac{2\left(1 +3 \alpha\right)\,T\,F'(T)}{\left(1 + \alpha\right)\,\left(\frac{F(T)}{2}-T\,F'(T)\right)}\right]\,\frac{K}{T}}{1+\left[\frac{6\,T\, F''(T)}{\left(\frac{F'(T)}{2}+T\,F''(T)\right)}\right]\,\frac{K}{T}}\right] 
\nonumber\\   
\end{eqnarray}
where $K=K(T)=\frac{\left|k\right|}{a^2}$.
If $\frac{K}{T}\ll 1$, we obtain:
\begin{align}\label{105}
\frac{dT}{d\tau}\approx &\,\mathcal{F}(T)\,-\,\mathcal{G}(T)\,K(T)
\end{align}

\subsubsection{Comment}

Note that in both cases, we have that:
\begin{eqnarray}\label{106}
\frac{dT}{d\tau} = \mathcal{F}(T)\,-k\,\mathcal{G}(T)\,K(T)
\end{eqnarray}
where $k=-1$ gives exactly the eqn. \eqref{102} with $K(T)=\frac{\sqrt{-k}}{a^2}$ and $k=+1$ gives exactly the eqn. \eqref{105} with $K(T)=\frac{\left|k\right|}{a^2}$. The relationship is exact when $k=0$. Clearly late time stability of the $k=0$ solutions to non-zero $k$ perturbations will occur when $\mathcal{F}(T) \gg \mathcal{G}(T)\,K(T)$ for $\tau\,\rightarrow\,\infty$.

\section{Quadratic $F(T)$ Solutions}

\noindent For example, if we have the theory $F(T) = -\Lambda +T+ \gamma\,T^{\beta}$, the functions $\mathcal{F}(T)$ and $\mathcal{G}(T)$ become, respectively:
\begin{eqnarray}
\mathcal{F}(T) &=& -\frac{3\,\left(1 + \alpha\right)\,\left[\Lambda +T+\gamma\,\left(2\beta-1\right)\,T^{\beta}\right]}{\left[1+\gamma\,\beta\,(2\beta-1)\,T^{\beta-1}\right]} ,    \label{107a}
\\
\mathcal{G}(T) &=& \frac{12\left(1 +3 \alpha\right)\,\left(1+\gamma\,\beta\,T^{\beta-1}\right)}{\left[1+\gamma\,\beta\,(2\beta-1)\,T^{\beta-1}\right]}
\nonumber\\
& &\quad\quad+\frac{9\,\left(1 + \alpha\right)\left[1-\gamma\,\beta\,(2\beta-3)\,T^{\beta-1}\right]\,\left[\Lambda +T+ \gamma\,(2\,\beta-1)\,T^{\beta}\right]}{T\,\left[1+\gamma\,\beta\,(2\beta-1)\,T^{\beta-1}\right]^2}.   \label{107b}
\end{eqnarray}
For $F(T) = -\Lambda +T+ \gamma\,T^{\beta}$, the $k=0$ solution (of eqn. \eqref{18b}) is given by:
\begin{eqnarray}\label{111a}
A_0\,\exp\left(-3\,\left(1 + \alpha\right)\tau\right) = \Lambda +T(\tau)+\gamma\,\left(2\beta-1\right)\,T^{\beta}(\tau),
\end{eqnarray}
where $A_0=2\tilde{A}_0$. In the quadratic case $\beta=2$ (explicit solutions in the cubic case $\beta=3$ are also possible), we then obtain from  \eqref{quad} the function $T(\tau)$ explicitly as:
\begin{eqnarray}\label{111d}
T(\tau) &=& \frac{1}{6\,\gamma}\,\left[-1 + \sqrt{1-12\gamma\,\left(\Lambda-A_0\,\exp\left(-3\,\left(1 + \alpha\right)\tau\right)\right)}\right],
\end{eqnarray} 
since $T>0$ (and $1-12\gamma\,\Lambda > 0$) and where 
\begin{align}\label{112}
H(\tau)=& \sqrt{\frac{T(\tau)}{6}}.
\end{align}
Note that as $\tau \rightarrow \infty$, $H(\tau)\rightarrow H_c$ and $T(\tau)\rightarrow\,T_c$, both positive constants for $\Lambda\neq 0$. From $\frac{d\tau}{dt}=H(\tau)$, we get by integration $\tau=\tau(t)$, or using $H=\frac{\dot{a}}{a}$, $\tau=\tau_0+\ln\left(a(t)\right)$. From the conservation eqns. $\rho(t)=\rho_0\,a^{-3\left(1+\alpha\right)\,t}$, so that $\rho\rightarrow 0$ as $t\rightarrow \infty$.

In the quadratic case eqns. \eqref{107a} and \eqref{107b} explicitly become:
\small
\begin{subequations}
\begin{eqnarray}
{\mathcal{F}(\tau)} &=& {-\frac{3\,\left(1 + \alpha\right)\,A_0\,\exp\left(-3\,\left(1 + \alpha\right)\tau\right)}{\sqrt{1-12\gamma\,\left(\Lambda-A_0\,\exp\left(-3\,\left(1 + \alpha\right)\tau\right)\right)}} , }
 \label{120a}
\\
{\mathcal{G}(\tau)} &=& {\frac{4\left(1 +3 \alpha\right)\,\left(2 + \sqrt{1-12\gamma\,\left(\Lambda-A_0\,\exp\left(-3\,\left(1 + \alpha\right)\tau\right)\right)}\right)}{\sqrt{1-12\gamma\,\left(\Lambda-A_0\,\exp\left(-3\,\left(1 + \alpha\right)\tau\right)\right)}}   }
\nonumber\\
& &\quad{ +\frac{18\gamma\,\left(1 + \alpha\right)\left[4 - \sqrt{1-12\gamma\,\left(\Lambda-A_0\,\exp\left(-3\,\left(1 + \alpha\right)\tau\right)\right)}\right]\,A_0\,\exp\left(-3\,\left(1 + \alpha\right)\tau\right)}{\left[-1 +  \sqrt{1-12\gamma\,\left(\Lambda-A_0\,\exp\left(-3\,\left(1 + \alpha\right)\tau\right)\right)}\right]\,\left[1-12\gamma\,\left(\Lambda-A_0\,\exp\left(-3\,\left(1 + \alpha\right)\tau\right)\right)\right]}.  }
\nonumber\\ \label{120b}
\end{eqnarray}
\end{subequations}
\normalsize
Alternatively, the eqns. \eqref{107a} and \eqref{107b} can be written for $\beta=2$ as:
\small
\begin{subequations}
\begin{eqnarray}
{\mathcal{F}(T)} &=& {-\frac{3\,\left(1 + \alpha\right)\,\left(\Lambda + T+ 3\gamma\,T^2\right)}{\left(1+6\gamma\,T\right) } , }
 \label{122a}
\\
{\mathcal{G}(T)} &=& {\frac{12\left(1 +3 \alpha\right)\,\left(1+2\,\gamma\,T\right)}{\left(1+6\gamma\,T\right)}   }+{ \frac{9\,\left(1 + \alpha\right)\left(1-2\,\gamma\,T\right)\,\left(\Lambda + T+ 3\gamma\,T^2 \right)}{T\,\left(1+6\,\gamma\,T\right)^2},  }  \label{122b}
\end{eqnarray}
\end{subequations}
\normalsize
where $T$ is given exactly by eq \eqref{111d}.

If $K=K(\tau)=\frac{|k|}{a^2}=|k|\,e^{-2\tau}$, then $K(T)=K_0\,\left(\Lambda + T+ 3\gamma\,T^2\right)^{\frac{2}{3\,\left(1 + \alpha\right)}}$ where $K_0$ is a constant, and the product of eqn. \eqref{122b} with $K(T)$ is:
\small
\begin{eqnarray}
{K(T)\,\mathcal{G}(T)} &=& {\frac{12\,K_0\,\left(1 +3 \alpha\right)\,\left(1+2\,\gamma\,T\right)}{\left(1+6\gamma\,T\right)} \left(\Lambda + T+ 3\gamma\,T^2\right)^{\frac{2}{3\,\left(1 + \alpha\right)}}  }
\nonumber\\
& &\quad +{ \frac{9\,K_0\,\left(1 + \alpha\right)\left(1-2\,\gamma\,T\right)}{T\,\left(1+6\,\gamma\,T\right)^2} \,\left(\Lambda + T+ 3\gamma\,T^2 \right)^{1+\frac{2}{3\,\left(1 + \alpha\right)}}.  }\label{124}
\end{eqnarray}
\normalsize
Due to the $T^{-1}$ term in the denominator of the second term, $K(T)\,\mathcal{G}(T)$ always diverges when $T_c=0$. Thus, in the case $\Lambda = 0$ all zero curvature solutions are unstable to the inclusion of non-zero spatial curvature.

In the case $\Lambda\, \neq\,0$, eqns. \eqref{122a} and \eqref{124} become for $T_c$ sufficiently less that unity:
\small
\begin{subequations}
\begin{eqnarray}
{\mathcal{F}(T)} &\approx & {-3\,\left(1 + \alpha\right)\,T\,\left[1-3\gamma\,T \right]} \label{127}
\\
{K(T)\,\mathcal{G}(T)}  &\approx & {12\,K_0\,\left(1 +3 \alpha\right)\,T^{\frac{2}{3(1+\alpha)}}\left[1+2\,\gamma\,T\left(\frac{1}{(1+\alpha)}-2\right)+\gamma^2\,T^2\left(\frac{2}{(1+\alpha)^2}-\frac{11}{(1+\alpha)}+24\right)\right]}
\nonumber\\
& &+{ 9\,K_0\,\left(1 + \alpha\right)\,T^{\frac{2}{3(1+\alpha)}}\left[1+\gamma\,T\,\left(\frac{2}{(1+\alpha)}-11\right)+\gamma^2\,T^2\left(\frac{2}{(1+\alpha)^2}-\frac{25}{(1+\alpha)}+90\right)\right]  }   \label{128}
\nonumber\\
&\,\equiv\,& T^{\frac{2}{3(1+\alpha)}} \left[1+A_1\,T+A_2\,T^2\right]
\end{eqnarray}
\normalsize
\end{subequations}
For local stability, we consequently need $T^{\frac{2}{3(1+\alpha)}} \ll T$ for $T\,\rightarrow\,0$, so that $\frac{2}{3(1+\alpha)}>1$ and finally $-1 < \alpha < -\frac{1}{3}$, corresponding to an inflating perfect fluid eqn. of state.

\section{Power-Law solutions}

Let us work in terms of the coordinate $t$ instead of $\tau$ (i.e., $a=a(\tau)=e^{\tau}$). By assuming $a(t)=a_n\,t^n$, so that
\begin{eqnarray}\label{155}
H(t)= \frac{n}{t} ,\quad\quad T(t)= \frac{6\,n^2}{t^2} , \quad\quad  \frac{dT(t)}{dt} = -\frac{12\,n^2}{t^3} = -\frac{2\,T(t)}{t} ,
\end{eqnarray}
where we include negative values of $n$ (corresponding to contracting models for $t>0$, but negative values of $t$ are also possible). We then get from eqn. \eqref{18b} for the total energy density $\rho_{sum}$ by adding the matter density ($\rho(t)$) and a cosmological constant source ($\Lambda$):
\begin{eqnarray}
\frac{F}{2} -T\,F^\prime &=&  -\tilde{A}_0\,\left[a(t)\right]^{-3(1+\alpha)}-\frac{\Lambda}{2}   , \label{154}
\end{eqnarray}
where $\tilde{A}_0$ is a positive constant. We recall that $-1<\alpha \leq 1$.

By substituting eqns. \eqref{155} into eqn. \eqref{154} and using $t=\frac{\sqrt{6}\,n}{\sqrt{T}}$, we obtain:
\begin{eqnarray}\label{157}
F(T) &=& \frac{C}{1-3(1+\alpha)n}\,T^{\left[\frac{3(1+\alpha)n}{2}\right]} + B\,\sqrt{T} - \Lambda.
\end{eqnarray}
where $B$ and $C$ are non-zero constants and $n \neq \frac{1}{3(1+\alpha)}$. By substituting eqn. \eqref{157} into eqns. \eqref{18} and \eqref{103}, the functions $\mathcal{F}(T)$ and $\mathcal{G}(T)$ are given by:
\begin{subequations}
\begin{align}\label{201}
\mathcal{F}(T) = -\frac{2}{n}\,T+\frac{2\,\Lambda}{n\,C}\,T^{1-\frac{3(1+\alpha)\,n}{2}}
\end{align}
\small
\begin{align}\label{202}
\mathcal{G}(T) =& \frac{6\left[(1-3\alpha)\,n-3\right]}{n\,\left[1-3(1+\alpha)\,n\right]}-\frac{4B\,\left[(1+3\alpha)n+1\right]}{(1+\alpha)\,C\,n^2}\,T^{\left[1-3(1+\alpha)\,n\right]/2}
\nonumber\\
& -4\Lambda\left[\frac{9\,\left[(1+\alpha)\,n-1\right]}{2\,C\,n\,\left[1-3(1+\alpha)\,n\right]}\,T^{-3(1+\alpha)\,n/2}-\frac{B}{C^2\,n^2\,(1+\alpha)}\,T^{\left[1-6(1+\alpha)\,n\right]/2}\right].
\end{align}
\normalsize
\end{subequations}
But we need to compare $\mathcal{F}(T)$ with $K(T)\,\mathcal{G}(T)$ where $K(T)=\frac{\left|k\right|}{a^2(t)}=K_0\,T^n$ according to eqn. \eqref{155}. The product of eqn. \eqref{202} with 
$K(T)$ is:
\small
\begin{align}\label{203}
K(T)\,\mathcal{G}(T) =& \frac{6\,K_0\left[(1-3\alpha)\,n-3\right]}{n\,\left[1-3(1+\alpha)\,n\right]}\,T^n-\frac{4B\,K_0\,\left[(1+3\alpha)n+1\right]}{(1+\alpha)\,C\,n^2}\,T^{\left[1-(1+3\alpha)\,n\right]/2}
\nonumber\\
& -4\Lambda\,K_0\left[\frac{9\,\left[(1+\alpha)\,n-1\right]}{2\,C\,n\,\left[1-3(1+\alpha)\,n\right]}\,T^{-(1+3\alpha)\,n/2}-\frac{B}{C^2\,n^2\,(1+\alpha)}\,T^{\left[1-2(2+3\alpha)\,n\right]/2}\right] .
\end{align}
\normalsize

\subsection{Stability Conditions}

We wish to consider the stability of the $k=0$ solutions to curvature perturbations (see comments earlier). The limit of eqn. \eqref{201} for $T \rightarrow 0$ (i.e., $t\rightarrow\,\infty$) is:
\begin{align}\label{204}
\mathcal{F}(T) \,\sim\,-\frac{2}{n}\,T +\frac{2\,\Lambda}{n\,C}\,T^{1-\frac{3(1+\alpha)\,n}{2}} ,
\end{align}
where $n \neq 0$. For $n<0$ (contracting model), eqn. \eqref{204} becomes for $T\,\rightarrow 0$:
\begin{align}\label{205}
\mathcal{F}(T) \,\sim\,T .
\end{align}
For $n>0$ (expanding model), eqn. \eqref{204} becomes for $T\,\rightarrow 0$:
\begin{align}\label{206}
\mathcal{F}(T) \,\sim\,T^{1-\frac{3(1+\alpha)\,n}{2}} ,
\end{align}
where $n \leq \frac{2}{3(1+\alpha)}$ for a converging $\mathcal{F}(T)$ when $T\,\rightarrow 0$.

\subsubsection{General $n<0$ Case}

We assume that $\mathcal{F}(T) \,\sim\,T$ and that $B \neq 0$ and $\Lambda \neq 0$. For stability, we obtain the following criteria for the first term of eqn. \eqref{203}:
\begin{eqnarray}\label{210}
(1-3\alpha)\,n-3=0 \quad\Rightarrow \quad n=\frac{3}{(1-3\alpha)} < 0, 
\end{eqnarray}
so that $\frac{1}{3}<\alpha \leq 1$. The eqn. \eqref{203} becomes:
\small
\begin{align}\label{211}
K(T)\,\mathcal{G}(T) =& -\frac{8B\,K_0\,(1-3\alpha)\left(2+3\alpha\right)}{9(1+\alpha)\,C}\,T^{\frac{1}{2}-\frac{3(1+3\alpha)}{2(1-3\alpha)}}
\nonumber\\
& -4\Lambda\,K_0\left[-\frac{3\,(1-3\alpha)\left(1+3\alpha\right)}{4C\,\left(2+3\alpha\right)}\,T^{-\frac{3(1+3\alpha)}{2(1-3\alpha)}}-\frac{B(1-3\alpha)^2}{9C^2\,(1+\alpha)}\,T^{\frac{1}{2}-\frac{3(2+3\alpha)}{(1-3\alpha)}}\right] ,
\end{align}
\normalsize
where $\alpha \neq -\frac{2}{3}$ (and $\alpha > -1$).

For stability, we need to compare all the powers of terms in eqn. \eqref{211} with unity (the power of $T$). Therefore, we require that simultaneously:
\begin{subequations}
\begin{eqnarray}
\frac{1}{2}-\frac{3(1+3\alpha)}{2(1-3\alpha)} &>& 1  , \label{212a}
\\
-\frac{3(1+3\alpha)}{2(1-3\alpha)} &>& 1 , \label{212b}
\\
\frac{1}{2}-\frac{3(2+3\alpha)}{(1-3\alpha)} &>& 1 , \label{212c} 
\end{eqnarray}
\end{subequations}
and we obtain the condition $\frac{1}{3}<\alpha \leq 1$ (for $K(T)\,\mathcal{G}(T) \ll \mathcal{F}(T)$).

\subsubsection{General $n>0$ Case}

In the more physical expanding case for $B \neq 0$ and $\Lambda \neq 0$, $\mathcal{F}(T)\,\sim\,T^{1-\frac{3(1+\alpha)\,n}{2}} $ when $T\,\rightarrow 0$, and so we compare all powers of the terms in eqn. \eqref{203} with the power of $T^{1-\frac{3(1+\alpha)\,n}{2}}$:
\begin{subequations}
\begin{eqnarray}
n &>& 1-\frac{3(1+\alpha)\,n}{2} , \label{220a}
\\
\frac{1}{2}-\frac{(1+3\alpha)\,n}{2} &>& 1-\frac{3(1+\alpha)\,n}{2} ,  \label{220b}
\\
-\frac{(1+3\alpha)\,n}{2} &>& 1-\frac{3(1+\alpha)\,n}{2} , \label{220c}
\\
\frac{1}{2}-(2+3\alpha)\,n &>& 1-\frac{3(1+\alpha)\,n}{2} . \label{220d}
\end{eqnarray}
\end{subequations}
From eqns. \eqref{220a} to \eqref{220d}, we obtain that $-1<\alpha<-\frac{1}{3}$ as stability conditions (with $n>1$ for $-1<\alpha \leq -\frac{2}{3}$ and $n>-\frac{1}{1+3\alpha}$ for $-\frac{2}{3}<\alpha<-\frac{1}{3}$).

\subsubsection{Special cases}

\noindent \textbf{For $B=0$ ($\Lambda \neq 0$), we have as stability conditions:}
\begin{itemize}
\item For $n<0$: we only need to satisfy eqn. \eqref{212b} and we still obtain $\frac{1}{3}<\alpha \leq 1$ for stable solutions ($K(T)\,\mathcal{G}(T)\,\sim\,T^{-\frac{3(1+3\alpha)}{2(1-3\alpha)}}$).

\item For $n>0$: we get from eqns. \eqref{220a} and \eqref{220c} for $\mathcal{F}(T)\,\sim\,T^{1-\frac{3(1+\alpha)\,n}{2}} $:
\begin{subequations}
\begin{eqnarray}
n &>& 1-\frac{3(1+\alpha)\,n}{2} , \label{223a}
\\
-\frac{(1+3\alpha)\,n}{2} &>& 1-\frac{3(1+\alpha)\,n}{2} . \label{223c}
\end{eqnarray}
\end{subequations}
From eqns. \eqref{223a} and \eqref{223c}, we obtain that $-1 < \alpha \leq 1$ and $n>1$ for stable solutions where $K(T)\,\mathcal{G}(T)\,\sim T^{-(1+3\alpha)\,n/2}$.

\end{itemize}

\noindent \textbf{For $\Lambda=0$ ($B \neq 0$), we have as stability conditions:}
\begin{itemize}
\item For $n<0$: we only need to satisfy eqn. \eqref{212a} and we still find that $\frac{1}{3}<\alpha \leq 1$ and then $K(T)\,\mathcal{G}(T)\,\sim\,T^{\frac{1}{2}-\frac{3(1+3\alpha)}{2(1-3\alpha)}}$.

\item For $n>0$: we have that $\mathcal{F}(T)\,\sim\,T$ and from eqn. \eqref{203} we obtain the stability conditions:
\begin{subequations}
\begin{eqnarray}
n &>& 1 , \label{221a}
\\
\frac{1}{2}-\frac{(1+3\alpha)\,n}{2} &>& 1  . \label{221b}
\end{eqnarray}
\end{subequations}
From eqns. \eqref{221a} and \eqref{221b}, we have that $-1<\alpha < -\frac{1}{3}$  for stable solutions (with $n>1$ for $-1<\alpha \leq -\frac{2}{3}$ and $n>-\frac{1}{1+3\alpha}$ for $-\frac{2}{3}<\alpha<-\frac{1}{3}$). These exotic fields with negative pressure are as appear in GR. Thus we get that $K(T)\,\mathcal{G}(T)\, \equiv \,A_1\,T^n+B_1\,T^{\left[1-(1+3\alpha)\,n\right]/2}$, and then:
\begin{subequations}
\begin{eqnarray}
K(T)\,\mathcal{G}(T)\,&\sim &\,T^n \quad\quad\quad\quad\quad\quad\; \text{if}\quad n < \frac{1}{3(1+\alpha)} , \label{221c}
\\
&\sim &\,T^{\left[1-(1+3\alpha)\,n\right]/2} \quad\; \text{if}\quad n > \frac{1}{3(1+\alpha)}.  \label{221d}
\end{eqnarray}
\end{subequations}

\end{itemize}

\noindent In the case $\Lambda=0$ and $B=0$, we obtain the GR solutions.

\subsubsection{Generalizations and physical consequences}

Physically we are primarily interested here in the stability of the $k=0$ TRW models to non-zero $k$ perturbations and the question of late time acceleration. The assumed eqn. of state is $p=\alpha\,\rho$, where $\alpha=\frac{1}{3}$ and $\alpha=0$ correspond to the physically important cases of radiation and dust (matter domination) ($\alpha=-1$ formally corresponds to a cosmological constant). However, more general perfect fluid sources with non-constant $\alpha$ are possible, as are even more general sources such as, for example, a non-interacting mixture of dust and radiation with $\rho=\rho_m + \rho_r$ and $p=\frac{\rho_r}{3}$, which also admits power-laws solutions. However, the important asymptotic behaviour of these models (e.g., the non-interacting two-fluid models alluded to above) will often be described by the $\alpha$ constant power-laws solutions above (hence illustrating the role of such constant $\alpha$ power law solutions for more physical cosmological models). Studying more physical models with a scalar field will need to be done using different techniques. A scalar field can be included with an effective $\rho_{\phi}$ and $p_{\phi}$, defined by:
\begin{subequations}
\begin{eqnarray}
\rho_{\phi} &=& \frac{\dot{\phi}^2}{2}+V(\phi) , \label{densityphi}
\\
p_{\phi} &=& \frac{\dot{\phi}^2}{2}-V(\phi) , \label{pressurephi}
\end{eqnarray}
\end{subequations}
subject to the corresponding conservation law which can be written in the form of the Klein-Gordon eqn:
\begin{align}\label{kleingordon}
0= \ddot{\phi}+V'(\phi)+3\,H\,\dot{\phi}, 
\end{align}
where $\phi$ is the scalar field, $V(\phi)$ is the scalar field potential and then $\dot{\phi}$  and $\ddot{\phi}$ are, respectively, the first and second time-derivatives of $\phi$ (all of these eqns are for a constant coupling function in $\phi$ \cite{darkenergy2}). The form of the potential $V(\phi)$ must have a physical motivation, and not all potentials will admit exact power-law solutions although solutions will often be power-law asymptotically.

\section{Discussion}

We have presented all teleparallel geometries that are invariant under the full $G_6$ Lie algebra of RW affine symmetries. We have discussed and clarified their properties, especially in the cases of non-zero $k$ (i.e., $k=-1$ and $k=+1$). In particular, we have explicitly computed the geometries in the proper coframes. We have displayed the correct FE for the $F(T)$ class of teleparallel TRW spacetimes.

In order to analyse the resulting cosmological models, we first considered the $k=0$ model and wrote the governing FE as an ordinary differential eqn., and discussed its properties. Subsequently, we formulated the $k \neq 0$ TRW models as pertubations of the $k=0$ model, which enabled us to carry out a late time stability analysis. In particular, we presented a detailed stability analysis for the quadratic $F(T) = -\Lambda + T + \gamma T^{2}$ models and the class of power-laws solutions, in both the case of a vanishing and non-vanishing $\Lambda$.

We explicitly considered models with a constant eqn. of state parameter $\alpha$, but made some comments on more general models with non-constant $\alpha$, such as models with a mixture of two fluids and scalar field models. It is the latter models, which include a scalar field, that are of more interest from a physical point of view (e.g., inflation). But different techniques are necessary for their investigation. We shall return to this in future work. Ultimately, of course, the models need to be confronted with observations \cite{Bahamonde:2021gfp,saridakis2} (see for example \cite{saridakis}).

\vspace{6pt} 



\section*{Acknowledgments}

AAC is funded by the Natural Sciences and Engineering Research Council of Canada. AL is funded by an AARMS fellowship. We would like to thank Robert van den Hoogen and David McNutt for helpful comments.




\section*{Abbreviations}
The following abbreviations are used in this manuscript:\\

\noindent 
\begin{tabular}{@{}ll}
DE & Differential Equation\\
FE & Field Equation\\
GR & General Relativity\\
KV & Killing Vectors\\
RW & Robertson-Walker\\
TRW  & Teleparallel Robertson-Walker
\end{tabular}





\end{document}